\pdfoutput=1
\documentclass[aps,pra,showkeys,showpacs,notitlepage,superscriptaddress]{revtex4-1}

\usepackage{amsmath}
\usepackage{amssymb}
\usepackage{hyperref}
\usepackage{graphics}
\usepackage{subfigure}
\usepackage{enumerate}
\usepackage{xspace}
\usepackage{fullpage}
\usepackage{hyperref}
\usepackage{tikz}
\usepackage{color}
\usetikzlibrary{arrows.meta, shapes, calc}

\newcommand{\Tr}{\mathrm{Tr}}
\newcommand{\tr}{\Tr}

\newcommand{\ket}[1]{\ensuremath{|#1\rangle}}
\newcommand{\bra}[1]{\ensuremath{\langle#1|}}
\newcommand{\ketbra}[2]{\ensuremath{\ket{#1}\bra{#2}}}

\newcommand{\1}{{\rm 1\hspace{-0.9mm}l}}
\newcommand{\Id}{\1}
\newcommand{\ii}{\mathrm{i}}
\newcommand{\dd}{\mathrm{d}}

\newcommand{\Sx}{\sigma_x}
\newcommand{\Sz}{\sigma_z}
\newcommand{\Sy}{\sigma_y}

\newcommand{\ie}{\emph{i.e.}\xspace}

\newcommand{\qutip}{QuTIP\xspace}

\newcommand{\tensorflow}{\textsc{TensorFlow}\xspace}

\begin{document}

%%%%%%%%%%%%%%%%%%%%%%%%%%%%%%%%%%%%%%%%%%%%%%%%%%%%%%%%%%%%%%%%%%%%%%%%%%%%%%%%
\title{Approximation of quantum control correction scheme using deep neural 
networks}

\author{M. Ostaszewski}
\affiliation{Institute of Theoretical and Applied Informatics, Polish Academy of
  Sciences, Ba{\l}tycka 5, 44-100 Gliwice, Poland}
\author{J.A. Miszczak}
\affiliation{Institute of Theoretical and Applied Informatics, Polish Academy of
    Sciences, Ba{\l}tycka 5, 44-100 Gliwice, Poland}
\author{L. Banchi}
\affiliation{QOLS, Blackett Laboratory, Imperial College London, SW7 2AZ, UK}
\author{P. Sadowski}
\affiliation{Institute of Theoretical and Applied Informatics, Polish Academy of
Sciences, Ba{\l}tycka 5, 44-100 Gliwice, Poland}

%\date{v. 1.10 (21/12/2018)}

\begin{abstract}
We study the functional relationship between quantum control pulses in the
idealized case and the pulses in the presence of an unwanted drift. We show that
a class of artificial neural networks called LSTM is able to model this
functional relationship with high efficiency, and hence the correction scheme
required to counterbalance the effect of the drift. Our solution allows studying
the mapping from quantum control pulses to system dynamics and analysing its behaviour with respect to the local variations in the control profile.
\end{abstract}



\maketitle

%%%%%%%%%%%%%%%%%%%%%%%%%%%%%%%%%%%%%%%%%%%%%%%%%%%%%%%%%%%%%%%%%%%%%%%%%%%%%%%%
\section{Introduction}
%%%%%%%%%%%%%%%%%%%%%%%%%%%%%%%%%%%%%%%%%%%%%%%%%%%%%%%%%%%%%%%%%%%%%%%%%%%%%%%%
The main objective and motivation of quantum information processing is the
development of new technologies based on principles of quantum mechanics such as
superposition and entanglement~\cite{dowling2003quantum}. Quantum technologies
require the development of methods and principles of quantum control, the
control theory of the quantum mechanical
system~\cite{dalessandro2007introduction}. Such methods have to be developed by
taking into account the behaviour of quantum systems~\cite{Gough2013, Pawela2014}. In particular, as quantum
systems are very susceptible to noise, which may influence the results of the
computation, the methods of quantum control have to include the means for
counteracting the decoherence~\cite{viola1998dynamical}.

The presented work is focused on the development of tools suitable for analysing
the relation between the control pulses used for idealized quantum systems, and
the control pulses required to execute the quantum computation in the presence
of undesirable dynamics. Modelling this relation is important to better understand the manifold of
	control pulses in the presence of noise, a case which is still poorly
	understood.  We focus on quantum dynamics described by a quantum
spin chain. We are interested in a method of approximating the correction
function of normal control pulses (NCP), \ie the function accepting control
pulses corresponding to the system without the drift Hamiltonian and generating
the de-noising control pulses (DCP) for the system with the drift Hamiltonian.
The existence of this function is non-trivial, since there 
are infinitely many pulses that produce the same evolution.
We propose an approximation that not only incorporates the global features of
the function but also describes its local properties. This feature is in
contrast with available methods based on optimisation which do not take into
account the continuous behaviour of the map from NCP to the de-noising control
pulses. Indeed, without further assumptions, in view of the non-injectivity of 
	quantum control, we may expect no relation at all between control pulses obtained 
from the optimization of different problems. On the other hand, we show that machine 
learning methods can be used for this purpose.

Recently, significant research effort has been invested in the application of
machine learning methods in quantum information processing
\cite{ciliberto2018quantum,dunjko2017machine,ostaszewski2018geometrical}. In particular, optimization
techniques borrowed from machine learning have been used to optimize the
dynamics of quantum systems~\cite{van2018learning}, either for quantum control
\cite{zahedinejad2014evolutionary, august2018taking} and simulation \cite{las2016genetic}, or for
implementing quantum gates with suitable time-independent Hamiltonians
\cite{banchi16learning}. These techniques include also quantum control
techniques from dynamic optimization \cite{sridharan2008gate} and reinforcement
learning \cite{bukov2017machine,niu2018universal}. In the presence of noise, neural networks give tools for optimizing dynamical decoupling, which can be seen as a quantum control correction scheme as considered by us, in a special case of the target operation being identity~\cite{august2017using}. On the level of gate decomposition, neural
networks have also been applied to the problem of decomposing arbitrary
operations as a sequence of elementary gate sets~\cite{swaddle17generating,fosel2018reinforcement}.

In this paper, we propose a method, based on an artificial neural network (ANN),
to study the correction scheme between control pulses obtained in
the ideal case, and those obtained when the system is subject to undesired dynamics. Moreover, we demonstrate that the utilized network has high
efficiency and can be used to analysis of the properties of the model.

This paper is organized as follows. In Section~\ref{sec:preliminaries} we
introduce the model and describe the architecture of a deep neural network which
will be used as an approximation function. In Section~\ref{sec:results} we
demonstrate that the proposed methods can be used for generating control pulses
without the explicit information about the model of a quantum system. We also
utilize it for the purpose of analysing the properties of the correction scheme.
Section \ref{sec:final} contains summary of the presented results.

%%%%%%%%%%%%%%%%%%%%%%%%%%%%%%%%%%%%%%%%%%%%%%%%%%%%%%%%%%%%%%%%%%%%%%%%%%%%%%%%
\section{Methods: Model and solution}\label{sec:preliminaries}
%%%%%%%%%%%%%%%%%%%%%%%%%%%%%%%%%%%%%%%%%%%%%%%%%%%%%%%%%%%%%%%%%%%%%%%%%%%%%%%%

In this section, we provide necessary notation and background information. We
start by introducing a spin-chain model and describe the problem of generating
quantum control pulses that counteract the undesired dynamics present in the
system. We also introduce the architecture of the artificial neural network used
to approximate the correction scheme.

%%%%%%%%%%%%%%%%%%%%%%%%%%%%%%%%%%%%%%%%%%%%%%%%%%%%%%%%%%%%%%%%%%%%%%%%%%%%%%%%
\subsection{Model of quantum system}
%%%%%%%%%%%%%%%%%%%%%%%%%%%%%%%%%%%%%%%%%%%%%%%%%%%%%%%%%%%%%%%%%%%%%%%%%%%%%%%%
Let us consider a system of two interacting qubits. The evolution of the system
is described by GKSL master equation 
\begin{equation}
\frac{\dd\rho}{\dd t}=-\ii[H(t),\rho] + 
\sum_j\gamma_j (L_j\rho L_j^\dagger - 
\frac{1}{2}\{L_j^\dagger L_j,\rho\}).
%\frac{\dd\ket{\psi}}{\dd t} =-\ii H(t)\ket{\psi}
\label{eq:schrod}
\end{equation}
where the Hamiltonian has three components
\begin{equation}
H(t) = H_c(t) + H_0+ H_{\gamma}.\label{eqn:model-hamiltonian}
\end{equation}
We consider the control Hamiltonian of the form
\begin{equation}
H_c(t) = h_x(t)\Sx\otimes\Id+ h_z(t)\Sz\otimes\Id, 
\label{eq:control_hamiltonian}
\end{equation}
and the base Hamiltonian 
\begin{equation}
H_0 = \Sx\otimes\Sx+\Sy\otimes\Sy+\Sz\otimes\Sz.
\end{equation}
The last element in Eq.~(\ref{eqn:model-hamiltonian}) is the drift Hamiltonian,
which can be an arbitrary two-qubit Hamiltonian multiplied by real
parameter $\gamma>0$. Incoherent part of Eq.~\ref{eq:schrod} models interaction with an environment with strengths $y_j$. It should be noted that in this paper will never consider
	the case when $\gamma$ and $\gamma_j$ both are different form zero.

Quantum optimal control refers to the task of executing a given unitary 
operation via the evolution of the system, 
in our case  described by Eq.~\eqref{eqn:model-hamiltonian}. 
 To achieve this, one has to properly choose 
the coefficients $h(t) = (h_x(t),h_z(t))$ in Eq.~\eqref{eq:control_hamiltonian}.
The set of reachable unitaries can be characterized \cite{dalessandro2007introduction} 
by studying the Lie algebra generated by the terms 
in Eq.~\eqref{eqn:model-hamiltonian}. For $H_\gamma=0$ our system is 
{\it fully-controllable}, so any target can be obtained with suitable choice of $h(t)$,
with no restriction on the control pulses. 
We assume that function $h(t)$ is piecewise constant in time slots $\Delta t_i =
[t_i,t_{i+1}]$, which are the equal partition of evolution time interval $T
=\bigcup_{i=0}^{n-1}\Delta t_i$. We also assume that $h_{x}(t)$ and $h_{z}(t)$
have values from interval $[ -1,1]$.

Function $h(t)$ will be represented as vectors of values of $h_x(t)$ and $h_z(t)$. For the case
$\gamma = 0$, we say that it represents NCP--- \emph{normal control pulses}.
Alternatively, for $\gamma \neq 0$, we say that $h(t)$ represents 
DCP---\emph{de-noising control pulses}. Since $h(t)$ is piecewise constant, both NCP and
DCP have two indices, with first index corresponding to time slots
$\{0,\ldots,n-1\}$, and the second index corresponding to the direction 
$\{x,z\}$, namely
\begin{equation}
\begin{split}
NCP_{i,j} &= h_j(t),\ \gamma=0,\\
DCP_{i,j} &= h_j(t),\  \gamma\neq0,\\ 
\end{split}
\end{equation}
where
\begin{equation}
t\in [t_i, t_{i+1}]\  i \in \{0,\ldots ,n-1\},\ j\in\{x,z\}.
\end{equation}
It is worth noting that the mapping from the set of control operations to the
unitary operator is not injective. Namely the same unitary can be obtained using
different choices of $h(t)$. To study the relationship between NCP and DCP we
need to select the DCP which is more closely related to the NCP. Because of
continuity, we do that numerically by using the NCP as starting guess of the
DCP. The final optimal DCP is then found using a local optimisation around the
initial NCP.

The figure of merit in our problem is the fidelity distance between 
superoperators, defined as \cite{floether12robust}
%{\bf LB: the following seems more a fidelity between Liouvillian evolutions 
%	(as in the previous notes)? 
%	Otherwise both $X(T)$ and $Y$ should have dimension $N\times N$, so I guess 
%	the normalization factor should be $\propto N^{-1} $. 
%	Since we don't consider anymore Liouvilleans, can we simplify the following definition?
%}
\begin{equation}
F=1-F_{err},
\end{equation} 
with
\begin{equation}
F_{err} = \frac{1}{2N^2} (\tr(Y-X(T))^\dagger(Y-X(T))),\label{eq:fidelity-error}
\end{equation} 
where $N$ is the dimension of the system in question, $Y$ is superoperator of the fixed target operator,
and $X(T)$ is evolution superoperator of operator resulting from the numerical integration of Eq.~\eqref{eq:schrod}
with given controls. In particular, for a target unitary operator $U$, its 
superoperator 
$Y$ is given by the formula
\begin{equation}
Y=U\otimes \bar{U}.
\end{equation}
Superoperator $X(T)$ is obtained from the unitary operator resulting from the 
integration of the Eq.~\eqref{eq:schrod}.

%%%%%%%%%%%%%%%%%%%%%%%%%%%%%%%%%%%%%%%%%%%%%%%%%%%%%%%%%%%%%%%%%%%%%%%%%%%%%%%%
\subsection{Architecture of artificial neural network}
%%%%%%%%%%%%%%%%%%%%%%%%%%%%%%%%%%%%%%%%%%%%%%%%%%%%%%%%%%%%%%%%%%%%%%%%%%%%%%%%
The control pulses used to drive the quantum system with Hamiltonian from
Eq.~\ref{eq:control_hamiltonian} are formally a time series. This aspect
suggests that one may study their properties using methods from pattern
recognition and machine learning \cite{bishop1995neural,goodfellow2016deep} that
have been successfully applied to process data with similar characteristics. The
mapping from NCP to DCP share similar mathematical properties with that of
statistical machine translation \cite{koehn2009statistical}, a problem which is
successfully modelled with artificial neural networks (ANN)
\cite{bahdanau2014neural}. Because of this analogy, we use ANN as the
approximation function to learn the correction scheme for control pulses. A
trained artificial neural network will be used as a map from NCP to DCP
\begin{equation}
\textrm{ANN}(\textrm{NCP})=\textrm{nnDCP},
\label{eq:ANN}
\end{equation}
where nnDCP, neural network DCP, is an approximation of DCP obtained by using the
neural network.

Because of time series character of control sequences, we utilize bidirectional
long short-term memory (LSTM) networks~\cite{hochreiter1997long}. The long
short-term memory block is a special kind of recurrent neural network (RNN), a
type of neural network with directed cycles between units. These cycles allow 
the RNN to keep track of the inputs received during the former times. 
%Similarly to other RNN, LSTM networks can take into account hidden states from their history.
In other words, the output at given time depends not only on current input,
but also on the history of earlier inputs. This kind of neural networks is
applicable in situations with time series with long-range correlations, such as
in natural language processing where the next word depends on the previous
sentence but also on the context.

\begin{figure}[ht!]
  \begin{tikzpicture}[>=LaTeX,
  ->,shorten >=1pt,
  cell/.style={
    align=center,
    rectangle, 
    rounded corners=2mm, 
    draw
  },
  input/.style={
  },
  hidden/.style={
  },
  subcell/.style={
    align=center,
    rectangle, 
    %circle,
    draw,
    minimum width = 1.cm,
    minimum height = 0.7cm,
    rounded corners=1mm
  },
  arrow/.style={
    rounded corners=.2cm
  }
  ]
  
  \pgfmathsetmacro{\yg}{-1.4}
  \pgfmathsetmacro{\yh}{-2.4}
  \pgfmathsetmacro{\yi}{-3.0}
  \pgfmathsetmacro{\as}{0.15}
  
  \node [cell, minimum height =1.cm, minimum width=5cm] (cell) at (2,0){\textbf{cell state}} ;
  \node [subcell] (forget) at (0,\yg) {\textbf{forget}\\\textbf{gate}};
  \node [subcell] (input) at (2,\yg) {\textbf{input}\\\textbf{gate}};
  \node [subcell] (candid) at (4cm,\yg) {\textbf{candidate}\\\textbf{gate}};
  \node [subcell] (output) at (6cm,\yg) {\textbf{output}\\\textbf{gate}};
  
  \node [cell, minimum height =1.cm, minimum width=1cm] (hidden) at (6,0) 
  {\textbf{hidden}\\\textbf{state}};
  
  \node [hidden] (prevh0) at  (-2,\yh) {$s_{t-1}$};
  \node [hidden] (prevhf) at  (0,\yh) {};
  \node [hidden] (prevhi) at  (2,\yh) {};
  \node [hidden] (prevhc) at  (4,\yh) {};
  \node [hidden] (prevho) at  (6,\yh) {};
  
  \node [input] (prevct) at (-2,0) {$C_{t-1}$};
  
  \node [input] (prev0) at  (-2,\yi) {$x_{t}$};
  \node [input] (prevf) at  (0,\yi) {};
  \node [input] (previ) at  (2,\yi) {};
  \node [input] (prevc) at  (4,\yi) {};
  \node [input] (prevo) at  (6,\yi) {};
  
  \draw [arrow] (prevct) -- (cell);
  
  \draw [arrow] (prevh0 -| prevhf) ++ (-1.4,0) -| ($(forget.south)-(\as,0)$);
  \draw [arrow] (prevh0 -| prevhf) ++ (-1.4,0) -| ($(input.south)-(\as,0)$);
  \draw [arrow] (prevh0 -| prevhf) ++ (-1.4,0) -| ($(candid.south)-(\as,0)$);
  \draw [arrow] (prevh0 -| prevhf) ++ (-1.4,0) -| ($(output.south)-(\as,0)$);
  
  \draw [arrow] (prev0 -| prevf) ++ (-1.4,0) -| ($(forget.south)+(\as,0)$);
  \draw [arrow] (prev0 -| prevf) ++ (-1.4,0) -| ($(input.south)+(\as,0)$);
  \draw [arrow] (prev0 -| prevf) ++ (-1.4,0) -| ($(candid.south)+(\as,0)$);
  \draw [arrow] (prev0 -| prevf) ++ (-1.4,0) -| ($(output.south)+(\as,0)$);
  
  \draw [arrow] (forget) -- ($(cell.south)-(2,0)$);;
  \draw [arrow] (input) -- (cell);
  \draw [arrow] (candid) -- ($(cell.south)+(2,0)$);
  \draw [arrow] (output) -- (hidden);
  
  \node [hidden] (outc) at (5,1.1) {$C_t$};
  \node [hidden] (outh) at (6,1.1) {$s_t$};
  
  \draw [arrow] ($(cell.east)-(0,\as)$) -- ($(hidden.west)-(0,\as)$);
  \draw [arrow] ($(cell.east)+(0,\as)$ -| hidden) ++ (cell) -| (outc);
  
  \draw [arrow] (hidden) -- (outh);
  
  \draw [arrow] (outc) -- (5,1.5) [rounded corners] -- (2.4,1.5);
  \draw [arrow] (outh) -- (6,1.9) [rounded corners] -- (2.4,1.9);;
  
  \node (dots1) at (2.0,1.5) {$\dots$};
  \node (dots2) at (2.0,1.9) {$\dots$};
  
  \draw [arrow] (1.5,1.5) -- (-3.0,1.5) [rounded corners] -- (-3.0,1) -- 
  (-3.0,0) -- (prevct) [rounded 
  corners];
  
  \draw [arrow] (1.5,1.9) -- (-3.4,1.9) [rounded corners] -- (-3.4,\yh)  -- 
  (prevh0) [rounded corners];

  \end{tikzpicture}
  \caption{Structure of LSTM network used in the experiments.}
\end{figure}
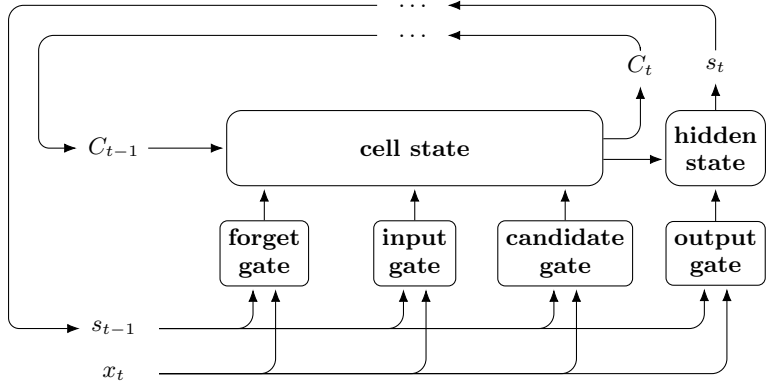

LSTM block consists of two cells
\begin{description}
  \item[cell state] 
  $
  c_t = c_{t-1}\circ \textrm{f}_t + \tilde{c}_t\circ \textrm{i}_t,
  $
  \item[hidden state]
  $
  s_t=\tanh (c_t)\circ \textrm{o}_t,
  $
\end{description}
which are constructed from following gates

\begin{description}
  \item[input gate]
  $
  \textrm{i}_t = \textrm{sigma}(x_tV^\textrm{i}+s_{t-1}W^\textrm{i}),
  $
  \item[forget gate]
  $
  \textrm{f}_t =\textrm{sigma}(x_tV^\textrm{f}+s_{t-1}W^\textrm{f}),
  $
  \item[output gate]
  $
  \textrm{o}_t = \textrm{sigma}(x_tV^\textrm{o}+s_{t-1}W^\textrm{o}),
  $
  \item[candidate]
  $
  \tilde{c}_t = \tanh(x_tV^\textrm{o}+s_{t-1}W^\textrm{o}),
  $
\end{description}
where $\circ$ is element wise multiplication of two vectors, $s_{t-1}$ is 
previous hidden state, and $x_t$ is an actual input state. Matrices $V$ and $W$ 
are the weight matrices of each gate. As one can see there is only one gate 
with hyperbolic tangent as activation function -- candidate gate. Rest of the 
gates has sigmoid activation function which has values from  $[0,1]$ interval. 
Using this function, neural networks decides which values 
are worth to keep and which should be forgotten. This gates maintain the memory 
of the network.

Basic architectures of RNN are not suitable for maintaining long time
dependences, due to the so-called \emph{vanishing/exploding gradient problem} --
the gradient of the cost function may either exponentially decay or explode as a
function of the hidden units. Thanks to the structure of LSTM the exploding gradient problem is reduced. The bidirectional version of LSTM
is characterized by the fact that it analyses the input sequence/time series
forwards and backwards. Thanks to this it uses not only information from the
past but also from the
future~\cite{schuster1997bidirectional,graves2005bidirectional}. For this
purpose the bidirectional LSTM unit consists of two LSTM units. One of them
takes as an input vector $[h(t_1), h(t_2),\ldots ,h(t_n)]$, and the second takes
$[h(t_{n}), h(t_{n-1}),\ldots ,h(t_1)]$.

As in Eq.~\eqref{eq:ANN}, the result of the network is a vector of control pulses 
nnDCP. To evaluate the quality of nnDCP we apply loss function described in 
Eq.~\ref{eq:fidelity-error}. To achieve this we integrate a new superoperator 
$X(T)$ form nnDCP and measure how far it is from the target superoperator~$Y$.

For two-qubit systems, we found that three stacked bidirectional LSTM layers are
sufficient for obtaining the high value of fidelity. Moreover, at the end of the
network we use one dense layer which processes the output of stacked LSTM to
obtain our nnDCP. Experiments are performed using \tensorflow
library~\cite{abadi2016tensorflow,tensorflow}.

%%%%%%%%%%%%%%%%%%%%%%%%%%%%%%%%%%%%%%%%%%%%%%%%%%%%%%%%%%%%%%%%%%%%%%%%%%%%%%%%%
\section{Results: Experiments}\label{sec:results}
%%%%%%%%%%%%%%%%%%%%%%%%%%%%%%%%%%%%%%%%%%%%%%%%%%%%%%%%%%%%%%%%%%%%%%%%%%%%%%%%%

For the purpose of testing the proposed method we use a sample of Haar random 
unitary matrices~\cite{mezzadrigenerate,miszczak12generating}. 
Pseudo-uniform random unitaries can be obtained also in the quantum control setting 
via random functions $h(t)$, provided that the control time $T$ is 
long enough \cite{banchi2017driven}. The exact implementation of sampling random Haar unitary matrices is available at~\cite{qcontrol_lstm_approx}.
In our experiments we use \qutip 
\cite{qutip,qutip1,qutip2} to generate control pulses to
training and testing data. 
First, we construct the target operators 

\begin{equation}
U_{target} = U\otimes \Id,
\end{equation}
where $U\in \mathbf{U}(2)$ is a random matrix. Next, using \qutip we 
generate NCP corresponding to the target operators. In the case of  our setup, 
the parameters are fixed as 
follows:
\begin{itemize}
	\item time of evolution $T=6$,
	\item number of intervals $n=32$,
	\item control pulses in $[-1, 1]$.
\end{itemize}

We train the network using a subset of the generated pairs \{(NCP, $Y_{target}$)\}, where $Y_{target}$ is a target superoperator obtained from $U_{target}$. In the training process network takes NCP as input and generates the nnDCP. Next this nnDCP is an input to the loss function. For this purpose we construct superoperator $X(T)$ resulting from integration of Eq.~\eqref{eq:schrod} using nnDCP. Next, we calculate error function between $X(T)$ and $Y_{target}$ as in Eq.~\eqref{eq:fidelity-error}. In this section we analyse only coherent drift i.e. in Eq.\eqref{eq:schrod} $\gamma_j = 0$.

Source code implementing experiments described in this paper is available at~\cite{qcontrol_lstm_approx}.

%%%%%%%%%%%%%%%%%%%%%%%%%%%%%%%%%%%%%%%%%%%%%%%%%%%%%%%%%%%%%%%%%%%%%%%%%%%%%%%%
\subsection{Performance of the neural network}
%%%%%%%%%%%%%%%%%%%%%%%%%%%%%%%%%%%%%%%%%%%%%%%%%%%%%%%%%%%%%%%%%%%%%%%%%%%%%%%%

The first experiment is designed to analyse the efficiency of the trained
network in terms of fidelity error of generated nnDCP control pulses. Trained
LSTM have mean fidelity on the test set as presented in
Table~\ref{tab:lstm-mean-fidelity}. It should be noted that, despite the fact
that the mean fidelity on the test set is high, the trained network sometimes
has outlier results \ie it returns nnDCP which corresponds to the operator with
high fidelity error.

\begin{table}[h!]
	\centering
	\begin{tabular}{ l@{\hspace{0.42cm}}c c c c}\hline &
        \multicolumn{4}{c}{$\gamma$} \\\cline{2-5}
		$H_{\gamma}  $  & $0.2$    & $0.4$ 	  & $0.6$    & $0.8$ \\\hline
		$\gamma(\Sy\otimes\Id)$ 					          & $.96$ & $.92$ & $.95$ &$.96$\\
		$\gamma(0.2\Sx\otimes\Id+0.8\Sy\otimes\Id)$ & $.97$ & $.94$ & $.97$ & $.97$ \\
		$\gamma(0.5\Sx\otimes\Id+0.5\Sy\otimes\Id)$ & $.98$ & $.96$ & $.98$ & $.97$ \\
		$\gamma(0.8\Sx\otimes\Id+0.2\Sy\otimes\Id)$ &$.99$ & $.97$ & $.98$ & $.99$
		
	\end{tabular}
\caption{The average of mean fidelity score. For each Hamiltonian and each 
parameter $\gamma$, 10 
independent experiments have been performed. It should be noted that first two columns with results i.e. comlums corresponding to $\gamma = 0.2$ and $0.4$ are obtained from network which takes mini batches equal to 5 and 4000 training samples and 1000 testing samples. Next two columns i.e. $\gamma = 0.6$ and $0.8$ are obtained from network with mini batches 50 and 8000 training samples and 4000 testing samples.}
\label{tab:lstm-mean-fidelity}
\end{table}

\begin{table}[h!]
  \centering
  \begin{tabular}{ l@{\hspace{0.42cm}}c c c c}\hline &
    \multicolumn{4}{c}{$\gamma$} \\\cline{2-5}
    $H_{\gamma}  $   & $0.6$    & $0.8$ \\\hline
    $\gamma(\Sy\otimes\Sy)$ 					& $.98$ & $.90$ \\
    $\gamma(0.1\Sx\otimes\Sx+0.8\Sy\otimes\Sy+0.1\Sz\otimes\Sz)$& $.98$&$.98$ \\
    $\gamma(0.1\Sx\otimes\Sx+0.6\Sy\otimes\Sy+0.3\Sz\otimes\Sz)$& $.98$&$.98$ \\
    $\gamma(0.2\Sx\otimes\Sx+0.6\Sy\otimes\Sy+0.2\Sz\otimes\Sz)$&$.96$& $.97$\\
    $\gamma(0.3\Sx\otimes\Sx+0.6\Sy\otimes\Sy+0.1\Sz\otimes\Sz)$&$.98$&$.98$ \\
  \end{tabular}
  \caption{The average mean fidelity scores for spin chain drifts. This results are obtained for network with mini batch equal to 10 and 8000 training samples and 4000 testing samples.}
  \label{tab:lstm-mean-fidelity-spinChain-drift}
\end{table}

The performed experiments show that it is possible to train LSTM networks for a
given system with high efficiency. Results from Table I are obtained by trained artificial neural networks on
different kinds of drift, \ie $\alpha\sigma_x\otimes\Id
+(1-\alpha)\sigma_y\otimes\Id$ for $\alpha\in\{0,0.2,0.5,0.8\}$, with different
values of $\gamma$. The experiment with this kind of drift is representative
because $\Sy$ is orthogonal to the controls $\Sx,\Sz$. The performed experiment
shows that it is possible to train LSTM to have the efficiency on the test set
above $90\%$ for chosen gammas. Some of results from
Table~\ref{tab:lstm-mean-fidelity} have average score lower than $90\%$. This is
caused by the outlier cases, when network performs with very low efficiency.
Such efficiency for chosen gammas is sufficient to our goal, which is to study
the relations between the system and control pulses for relatively small
disturbances.

In Table~\ref{tab:lstm-mean-fidelity-spinChain-drift} we present average score 
for model with spin chain drift. As 
one can see, these results are much better than in Table 
\ref{tab:lstm-mean-fidelity}. This can be explained by 
the fact that this type of drift is similar to the base Hamiltonian $H_0$ 
which is also a spin chain.

Results from above tables may suggest the answer for non trivial questions: for what kind of drift, map between NCP and DCP is easier to learn. If training on the model with drift operating the same system as the control is easier than training on the model with drift similar to base Hamiltonian?
Obtained results suggest that, when the
drift is asymmetric with respect to the base Hamiltonian, the neural network 
requires larger mini batch sizes to achieve similar efficiency.

%Intuitively, it is easier to change the input 
%NCP when the drift is similar to the base Hamiltonian. In the case when the 
%drift is asymmetric with respect to the base Hamiltonian, the neural network 
%requires larger mini batch sizes to achieve similar efficiency.

As one can see, the efficiency of the artificial neural networks depends on the 
choice of the hyperparameters. In our case, for some values of parameter 
$\gamma$ one needs to use bigger batch size and the larger training set to 
obtain satisfactory results. 
This behavior is incomprehensible because, for $\gamma=0.6$ and batch size equal to 5, RNN has problems with convergence. However, by increasing the batch size we were able to improve the
performance. We would like to stress that our aim was 
to show that it is possible to obtain mean efficiency greater than 90\%, not to 
examine what is the best possible efficiency of the network.

%%%%%%%%%%%%%%%%%%%%%%%%%%%%%%%%%%%%%%%%%%%%%%%%%%%%%%%%%%%%%%%%%%%%%%%%%%%%%%%%
\subsection{Utilization of the approximation}
\label{sec:exhibited_features}
%%%%%%%%%%%%%%%%%%%%%%%%%%%%%%%%%%%%%%%%%%%%%%%%%%%%%%%%%%%%%%%%%%%%%%%%%%%%%%%%
%The main benefit of treating the trained LSTM network as a function is the
%possibility of analysing the behaviour of the correction scheme by perturbing the
%input signals. For this purpose we analyse the variation of the outputs
%according to small changes in the inputs.

 In this section, we show the results
of an experiment, which allows checking what is the behaviour of the ANN() 
function when we perform local disturbances on a set of random NCP, and we check
what is the deviation of the new nnDCP from the original DCP. The original DCP is obtained from GRAPE algorithm initialized by NCP. 

Let us suppose that we have a trained LSTM. The procedure of checking its
sensitivity on variations of $h_j(t)$, for $j\in \{x,z\}$ and $t$ in the $i$-th
time slot, is as follows.

\newcounter{stepno}
\begin{list}{\textbf{Step \arabic{stepno}}}{\usecounter{stepno}}
    \item\label{st:variation-step1}  Select a NCP vector from the testing set and
    generate the corresponding nnDCP. % , which will be reference DCP. LB: don't understand this
    
	\item\label{st:variation-step} If the fidelity between target operator and
	operator resulting from nnDCP is lower than $90\%$ return to \textbf{Step
    \ref{st:variation-step1}}.

    \item\label{st:variation-step3} Select $i\in\{0,\ldots ,n-1\}$, 
    $j\in\{x,z\}$, change $(i,j)$ coordinate of NCP by fixed $\varepsilon$ 
	\begin{itemize}
		\item if $\textrm{NCP}_{i,j} < 1-\varepsilon$, then $\textrm{NCP}_{i,j} +\varepsilon$,
	
		\item if $\textrm{NCP}_{i,j} > 1-\varepsilon$, then $\textrm{NCP}_{i,j} -\varepsilon$,
	\end{itemize}
	 and calculate new collection of $\textrm{nnDCP}$ vectors with elements
	 $\textrm{nnDCP}^{i,j}$, denoting that the element was obtained by 
	 perturbing $j$-th component at the $i$-th time slot.
     
    \item\label{st:variation-step4} If the fidelity between target operator and
	operator resulting from $\textrm{nnDCP}^{i,j}$ is lower
	than $90\%$ return to \textbf{Step \ref{st:variation-step3}}.
    
    \item\label{st:variation-step5} Calculate norm $l_2$ (Euclidean
    norm) of difference between nnDCP and $\textrm{nnDCP}^{i,j}$.
\end{list}

Because of outlier results of the network, there are additional conditions on
generated DCP in the above algorithm. Applying the above algorithm for each NCP
from the testing set we can obtain a sample of variations. Next, we can analyse
empirical distributions of these variations.

\begin{figure}
	\centering
	\includegraphics[width=1\columnwidth]{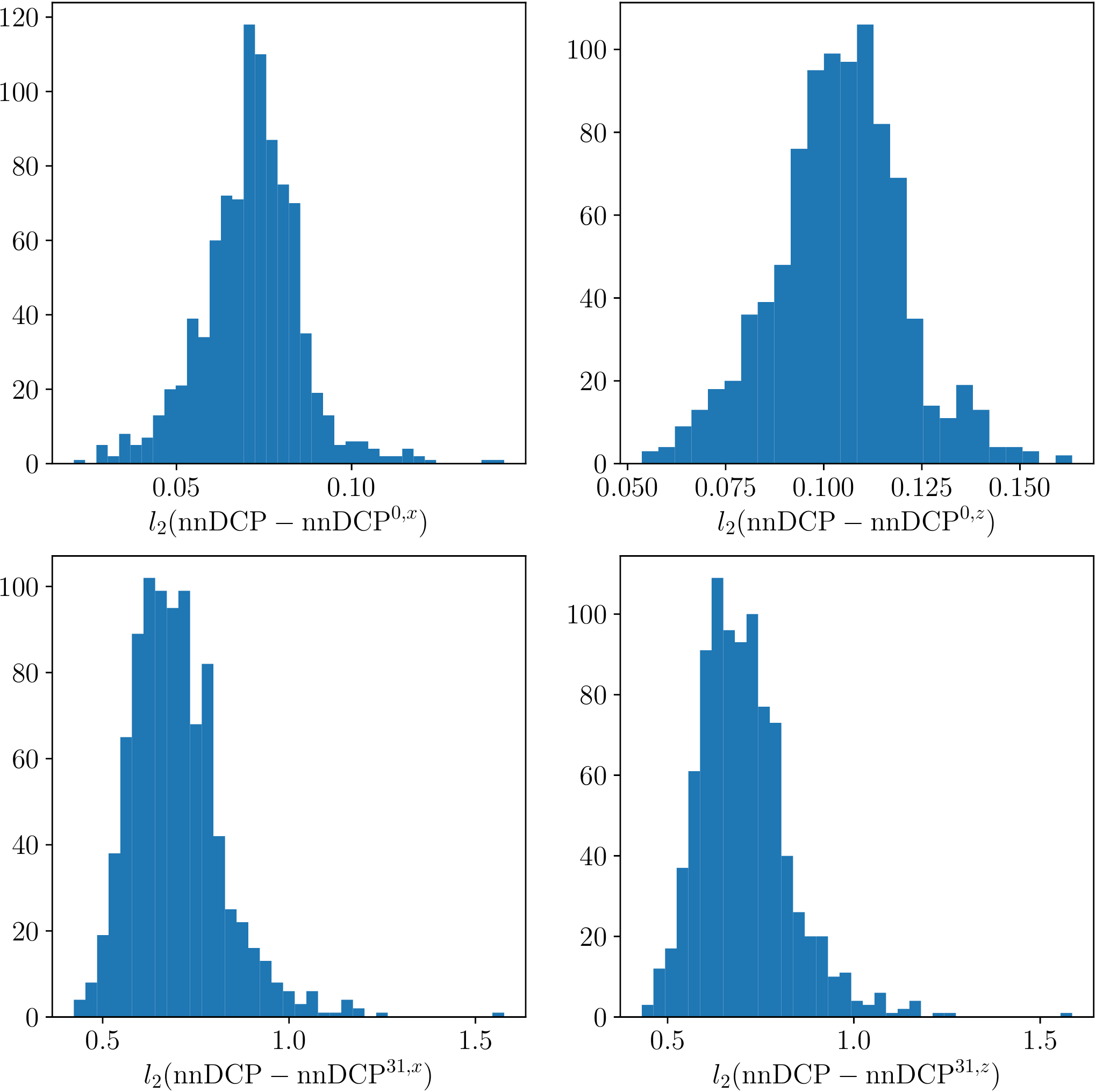}
	\caption{Exemplary histograms performing variation of ANN approximation. Histograms are generated from disturbances on set of 1000 NCP corresponding to random matrices, for fixed system $H_{\gamma} = \gamma \Sy\otimes\Id.$}
	\label{fig:hist_Sy_id}
\end{figure}

As an example of using the above method we consider the following example.
Let us suppose that the drift operator is of the form
\begin{equation}
H_{\gamma} = \gamma \Sy\otimes\Id.
\end{equation}
Then, for $\varepsilon =0.1$ and $\gamma=0.2$, the exemplary variation histograms are presented in Fig.~\ref{fig:hist_Sy_id}. Thanks to trained network we are able to analyse the impact of small changes of the input on the output (\textbf{Step~\ref{st:variation-step5}}). 
%As we can see, the interval of changes is similar for different controls ($h_x,\ h_z$), but differs for different time slots. 

\begin{figure}[ht]
	\subfigure[$H_{\gamma} = \gamma \Sy\otimes\Id$]{\includegraphics[]{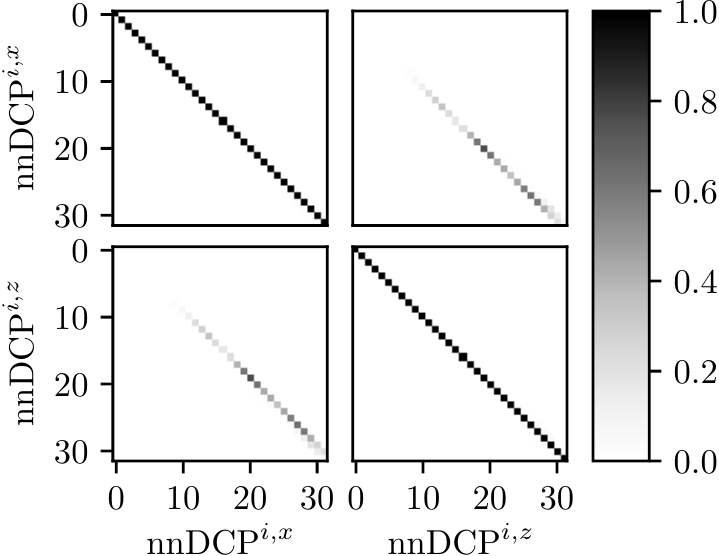}\label{fig:kruskal_matrix_hist_Sy_id}}	
	\subfigure[$H_{\gamma} = \gamma \Id\otimes\Sy$.]{\includegraphics[]{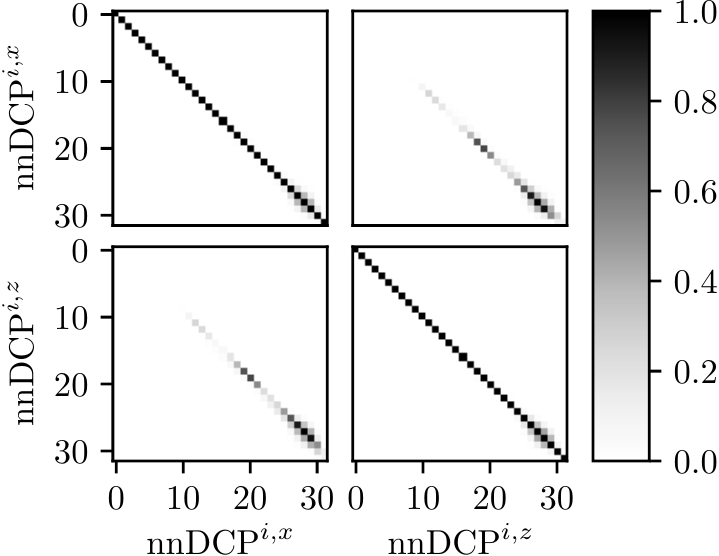}\label{fig:kruskal_matrix_hist_id_Sy}}
    \caption{Plots of $p$--values of Kruskal--Wallis test for tested drift
    Hamiltonians. Each element $(l,k)$ of above matrix plots represents a
    $p$--value of the test between empirical distributions of changes implied by
    disturbances on $l$-th time slot and $k$-th time slot. The horizontal axis
    of the left column and the vertical axis of top row plots correspond to
    disturbances on $\sigma_x$ control, while horizontal axis of the right
    column and the vertical axis of the bottom plots correspond to disturbances
    on $\sigma_x$ control.}
    \label{fig:kruskal_matrix_hist}
\end{figure}

One can consider the median of the distribution of variations as the
quantitative measure of the influence of the changes in the input signal on the
resulting DCP. To check if medians of distributions of changes are similar, we
perform Kruskal--Wallis statistical test for each pair of the changed
coordinates (see Fig.~\ref{fig:kruskal_matrix_hist_Sy_id}). Results presented
in~Fig~\ref{fig:kruskal_matrix_hist_Sy_id} shows that most of the distributions
received are statistically different regarding the median, \ie most of the
$p$--values is less than $0.05$. The values on the main diagonals, where we
compared disturbances on the same controls in the same time slots, are equal to
1. This observation confirms that the test behaves appropriately in this case.
On the other hand, one can observe that for time slots $10\leq i\leq30$
Kruskal--Wallis test for distributions obtained for $NCP_{i+1, x}\pm
\varepsilon$ and $NCP_{i,z}\pm \varepsilon$ gives $p$--values greater than
$0.05$. Thus, the disturbances introduced in this time slots on $h_x$ and $h_z$
coordinates of $NCP$ results in similar variations of the resulting DCP. The
situation is different for time slots $i\leq10$, where one can see that the
variation in DCP signal depends on which coordinate of the NCP signal is
disturbed.

The similar effect can be observed if the drift Hamiltonian acts on the second qubit only. In this case we can consider experiment where the drift Hamiltonian is of the form
\begin{equation}
H_{\gamma} = \gamma \Id\otimes\Sy, 
\end{equation}
with $\varepsilon =0.1$ and $\gamma =0.2$. 

The results of Kruskal--Wallis test for this situation are presented in
Fig.~\ref{fig:kruskal_matrix_hist_id_Sy}. One can see that our approach
suggests that there are similarities in distributions of variations implied by
disturbances on different controls and near time slots. This suggests that local
disturbances in control signals have a similar effect in the case of drift on
the target system and drift on the auxiliary system.

Moreover, one can see that the constructed approximation exhibits symmetry of
the model. This effect can be observed by analysing the disturbances of $h_x$
and $h_z$ controls in the same time slot. From the performed experiments one can
see they give similar variations quantified by the distribution of DCP changes.

One should note that on plots in Fig.~\ref{fig:kruskal_matrix_hist}, where we
compare variations of $\textrm{nnDCP}^{i,x}$ and $\textrm{nnDCP}^{i,z}$,
$p$--values greater than $0.05$ are focused along the diagonal. This symmetry is
not perfect, but one should note that training data are generated from random
unitary matrices. Moreover, as we do not impose any restrictions on the training
set to ensure the uniqueness of the correction scheme.

%%%%%%%%%%%%%%%%%%%%%%%%%%%%%%%%%%%%%%%%%%%%%%%%%%%%%%%%%%%%%%%%%%%%%%%%%%%%%%%%
\section{Discussions}\label{sec:discuss}
%%%%%%%%%%%%%%%%%%%%%%%%%%%%%%%%%%%%%%%%%%%%%%%%%%%%%%%%%%%%%%%%%%%%%%%%%%%%%%%%
In a case where Hamiltonian drift is absent $\gamma\neq 0$ and we want to find 
proper DCP for system with incoherent noise, things get more complicated. In 
our experiments we tested three kinds of Lindblad noise, namely
\begin{enumerate}
	\item with one Lindblad operator $L= \ketbra{0}{1}\otimes\Id$,
	\item with two Lindblad operators $L_1 = \sigma_z\otimes\Id$ and $L_2 = \Id\otimes\sigma_z$,
	\item and mixed two Lindblad operators $L_1 = \sigma_z\otimes\Id$ and $L_2= \ketbra{0}{1}\otimes\Id$.
\end{enumerate}
For all this open system noise models, nnDCP obtains efficiency above 90\% for 
$\gamma_1=\gamma_2\approx 0.01$. However, this strength of noise is so small 
that also NCP obtains similar efficiency. Therefore, the proposed model of the 
neural network does not correct NCP effectively, for incoherent types of noise.

Another important case in the context of quantum control is the robustness on the random fluctuations. In our experiments, we tried to generate nnDCP which will be robust for Gaussian fluctuations i.e. 
\begin{equation}
\forall i\in\{0,\ldots , n-1\},\ j\in\{x,z\}\  nnDCP_{i,j} +\delta_{i,j},
\end{equation}
where $\delta_{i,j} \sim \mathcal{N}(0,\sigma)$. The standard deviation we tested in two cases $\sigma =0.1$ and 0.2, while the model of Hmiltonian drift was $0.4(\Sy\otimes \Id)$.

 During the training process, returned by ANN control pulses were copied ten 
 times and to each copy we added Gaussian fluctuations. Such disturbed nnDCP 
 were applied to cost function Eq.~\eqref{eq:fidelity-error}, and next the 
 average fidelity was calculated. 
 Unfortunately, after this training process artificial neural network doesn't 
 produce nnDCP which are more robust for random fluctuations.  
 This failure might be caused by the fact that training of artificial neural 
 network is based on gradient descent. As we know from~\cite{niu2018universal}, 
 gradient based algorithm does not give very good results. On the other hand, task with which 
 we try to face is slightly different from the one in \cite{niu2018universal}, 
 because we want to generalize correction scheme over all target unitaries.

%%%%%%%%%%%%%%%%%%%%%%%%%%%%%%%%%%%%%%%%%%%%%%%%%%%%%%%%%%%%%%%%%%%%%%%%%%%%%%%%
\section{Concluding remarks}\label{sec:final}
%%%%%%%%%%%%%%%%%%%%%%%%%%%%%%%%%%%%%%%%%%%%%%%%%%%%%%%%%%%%%%%%%%%%%%%%%%%%%%%%

The primary objective of the presented work is to use artificial neural networks
for the purpose of approximating the structure of quantum systems. We propose to
use a bidirectional LSTM neural network. We argue that this type of artificial
neural network is suitable to capture time-dependences present in quantum
control pulses. We have developed a method of reconstructing the relation
between control pulses in idealised case and control pulses required to
implement quantum computation in the presence of undesirable dynamics. We argue
that the proposed method can be a useful tool to study the manifold of quantum control pulses in the noisy regime, 
	and to define new theoretical approaches to control noisy quantum systems.

\begin{acknowledgements}
LB acknowledges support from the UK EPSRC grant EP/K034480/1. MO acknowledges
support from Polish National Science Center scholarship 2018/28/T/ST6/00429. JAM
acknowledges support from Polish National Science Center grant
2014/15/B/ST6/05204. Authors would like to thank Daniel Burgarth for discussions
about quantum control, Bartosz Grabowski and Wojciech Masarczyk for discussions concerning the details of LSTM architecture, and Izabela Miszczak for reviewing the manuscript. Numerical calculations
were  possible thanks  to the support  of PL-Grid  Infrastructure.
\end{acknowledgements}

%%%%%%%%%%%%%%%%%%%%%%%%%%%%%%%%%%%%%%%%%%%%%%%%%%%%%%%%%%%%%%%%%%%%%%%%%%%%%%%%
\bibliography{lstm_aprox}

%merlin.mbs apsrev4-1.bst 2010-07-25 4.21a (PWD, AO, DPC) hacked
%Control: key (0)
%Control: author (8) initials jnrlst
%Control: editor formatted (1) identically to author
%Control: production of article title (-1) disabled
%Control: page (0) single
%Control: year (1) truncated
%Control: production of eprint (0) enabled
\begin{thebibliography}{36}%
\makeatletter
\providecommand \@ifxundefined [1]{%
 \@ifx{#1\undefined}
}%
\providecommand \@ifnum [1]{%
 \ifnum #1\expandafter \@firstoftwo
 \else \expandafter \@secondoftwo
 \fi
}%
\providecommand \@ifx [1]{%
 \ifx #1\expandafter \@firstoftwo
 \else \expandafter \@secondoftwo
 \fi
}%
\providecommand \natexlab [1]{#1}%
\providecommand \enquote  [1]{``#1''}%
\providecommand \bibnamefont  [1]{#1}%
\providecommand \bibfnamefont [1]{#1}%
\providecommand \citenamefont [1]{#1}%
\providecommand \href@noop [0]{\@secondoftwo}%
\providecommand \href [0]{\begingroup \@sanitize@url \@href}%
\providecommand \@href[1]{\@@startlink{#1}\@@href}%
\providecommand \@@href[1]{\endgroup#1\@@endlink}%
\providecommand \@sanitize@url [0]{\catcode `\\12\catcode `\$12\catcode
  `\&12\catcode `\#12\catcode `\^12\catcode `\_12\catcode `\%12\relax}%
\providecommand \@@startlink[1]{}%
\providecommand \@@endlink[0]{}%
\providecommand \url  [0]{\begingroup\@sanitize@url \@url }%
\providecommand \@url [1]{\endgroup\@href {#1}{\urlprefix }}%
\providecommand \urlprefix  [0]{URL }%
\providecommand \Eprint [0]{\href }%
\providecommand \doibase [0]{http://dx.doi.org/}%
\providecommand \selectlanguage [0]{\@gobble}%
\providecommand \bibinfo  [0]{\@secondoftwo}%
\providecommand \bibfield  [0]{\@secondoftwo}%
\providecommand \translation [1]{[#1]}%
\providecommand \BibitemOpen [0]{}%
\providecommand \bibitemStop [0]{}%
\providecommand \bibitemNoStop [0]{.\EOS\space}%
\providecommand \EOS [0]{\spacefactor3000\relax}%
\providecommand \BibitemShut  [1]{\csname bibitem#1\endcsname}%
\let\auto@bib@innerbib\@empty
%</preamble>
\bibitem [{\citenamefont {Dowling}\ and\ \citenamefont
  {Milburn}(2003)}]{dowling2003quantum}%
  \BibitemOpen
  \bibfield  {author} {\bibinfo {author} {\bibfnamefont {J.}~\bibnamefont
  {Dowling}}\ and\ \bibinfo {author} {\bibfnamefont {G.}~\bibnamefont
  {Milburn}},\ }\href {\doibase 10.1098/rsta.2003.1227} {\bibfield  {journal}
  {\bibinfo  {journal} {Phil. Trans. R. Soc. A}\ }\textbf {\bibinfo {volume}
  {361}},\ \bibinfo {pages} {1655} (\bibinfo {year} {2003})}\BibitemShut
  {NoStop}%
\bibitem [{\citenamefont {d'Alessandro}(2007)}]{dalessandro2007introduction}%
  \BibitemOpen
  \bibfield  {author} {\bibinfo {author} {\bibfnamefont {D.}~\bibnamefont
  {d'Alessandro}},\ }\href@noop {} {\emph {\bibinfo {title} {Introduction to
  quantum control and dynamics}}}\ (\bibinfo  {publisher} {CRC Press},\
  \bibinfo {year} {2007})\BibitemShut {NoStop}%
\bibitem [{\citenamefont {Gough}\ and\ \citenamefont
  {Belavkin}(2013)}]{Gough2013}%
  \BibitemOpen
  \bibfield  {author} {\bibinfo {author} {\bibfnamefont {J.~E.}\ \bibnamefont
  {Gough}}\ and\ \bibinfo {author} {\bibfnamefont {V.~P.}\ \bibnamefont
  {Belavkin}},\ }\href {\doibase 10.1007/s11128-012-0491-7} {\bibfield
  {journal} {\bibinfo  {journal} {Quantum Information Processing}\ }\textbf
  {\bibinfo {volume} {12}},\ \bibinfo {pages} {1397} (\bibinfo {year}
  {2013})}\BibitemShut {NoStop}%
\bibitem [{\citenamefont {Pawela}\ and\ \citenamefont
  {Pucha{\l}a}(2014)}]{Pawela2014}%
  \BibitemOpen
  \bibfield  {author} {\bibinfo {author} {\bibfnamefont {{\L}.}~\bibnamefont
  {Pawela}}\ and\ \bibinfo {author} {\bibfnamefont {Z.}~\bibnamefont
  {Pucha{\l}a}},\ }\href@noop {} {\bibfield  {journal} {\bibinfo  {journal}
  {Quantum Information Processing}\ }\textbf {\bibinfo {volume} {13}},\
  \bibinfo {pages} {227} (\bibinfo {year} {2014})}\BibitemShut {NoStop}%
\bibitem [{\citenamefont {Viola}\ and\ \citenamefont
  {Lloyd}(1998)}]{viola1998dynamical}%
  \BibitemOpen
  \bibfield  {author} {\bibinfo {author} {\bibfnamefont {L.}~\bibnamefont
  {Viola}}\ and\ \bibinfo {author} {\bibfnamefont {S.}~\bibnamefont {Lloyd}},\
  }\href {\doibase 10.1103/PhysRevA.58.2733} {\bibfield  {journal} {\bibinfo
  {journal} {Phys. Rev. A}\ }\textbf {\bibinfo {volume} {58}},\ \bibinfo
  {pages} {2733} (\bibinfo {year} {1998})}\BibitemShut {NoStop}%
\bibitem [{\citenamefont {Ciliberto}\ \emph {et~al.}(2018)\citenamefont
  {Ciliberto}, \citenamefont {Herbster}, \citenamefont {Ialongo}, \citenamefont
  {Pontil}, \citenamefont {Rocchetto}, \citenamefont {Severini},\ and\
  \citenamefont {Wossnig}}]{ciliberto2018quantum}%
  \BibitemOpen
  \bibfield  {author} {\bibinfo {author} {\bibfnamefont {C.}~\bibnamefont
  {Ciliberto}}, \bibinfo {author} {\bibfnamefont {M.}~\bibnamefont {Herbster}},
  \bibinfo {author} {\bibfnamefont {A.~D.}\ \bibnamefont {Ialongo}}, \bibinfo
  {author} {\bibfnamefont {M.}~\bibnamefont {Pontil}}, \bibinfo {author}
  {\bibfnamefont {A.}~\bibnamefont {Rocchetto}}, \bibinfo {author}
  {\bibfnamefont {S.}~\bibnamefont {Severini}}, \ and\ \bibinfo {author}
  {\bibfnamefont {L.}~\bibnamefont {Wossnig}},\ }in\ \href@noop {} {\emph
  {\bibinfo {booktitle} {Proc. R. Soc. A}}},\ Vol.\ \bibinfo {volume} {474}\
  (\bibinfo {organization} {The Royal Society},\ \bibinfo {year} {2018})\ p.\
  \bibinfo {pages} {20170551}\BibitemShut {NoStop}%
\bibitem [{\citenamefont {Dunjko}\ and\ \citenamefont
  {Briegel}(2017)}]{dunjko2017machine}%
  \BibitemOpen
  \bibfield  {author} {\bibinfo {author} {\bibfnamefont {V.}~\bibnamefont
  {Dunjko}}\ and\ \bibinfo {author} {\bibfnamefont {H.}~\bibnamefont
  {Briegel}},\ }\href {\doibase 10.1088/1361-6633/aab406} {\  (\bibinfo {year}
  {2017}),\ 10.1088/1361-6633/aab406},\ \Eprint
  {http://arxiv.org/abs/1709.02779} {arXiv:1709.02779} \BibitemShut {NoStop}%
\bibitem [{\citenamefont {Ostaszewski}\ \emph {et~al.}()\citenamefont
  {Ostaszewski}, \citenamefont {Miszczak},\ and\ \citenamefont
  {Sadowski}}]{ostaszewski2018geometrical}%
  \BibitemOpen
  \bibfield  {author} {\bibinfo {author} {\bibfnamefont {M.}~\bibnamefont
  {Ostaszewski}}, \bibinfo {author} {\bibfnamefont {J.}~\bibnamefont
  {Miszczak}}, \ and\ \bibinfo {author} {\bibfnamefont {P.}~\bibnamefont
  {Sadowski}},\ }\href@noop {} {\enquote {\bibinfo {title} {Geometrical versus
  time-series representation of data in learning quantum control},}\ }\Eprint
  {http://arxiv.org/abs/1803.05169} {arXiv:1803.05169} \BibitemShut {NoStop}%
\bibitem [{\citenamefont {van Nieuwenburg}\ \emph {et~al.}(2018)\citenamefont
  {van Nieuwenburg}, \citenamefont {Bairey},\ and\ \citenamefont
  {Refael}}]{van2018learning}%
  \BibitemOpen
  \bibfield  {author} {\bibinfo {author} {\bibfnamefont {E.}~\bibnamefont {van
  Nieuwenburg}}, \bibinfo {author} {\bibfnamefont {E.}~\bibnamefont {Bairey}},
  \ and\ \bibinfo {author} {\bibfnamefont {G.}~\bibnamefont {Refael}},\
  }\href@noop {} {\bibfield  {journal} {\bibinfo  {journal} {Physical Review
  B}\ }\textbf {\bibinfo {volume} {98}},\ \bibinfo {pages} {060301} (\bibinfo
  {year} {2018})}\BibitemShut {NoStop}%
\bibitem [{\citenamefont {Zahedinejad}\ \emph {et~al.}(2014)\citenamefont
  {Zahedinejad}, \citenamefont {Schirmer},\ and\ \citenamefont
  {Sanders}}]{zahedinejad2014evolutionary}%
  \BibitemOpen
  \bibfield  {author} {\bibinfo {author} {\bibfnamefont {E.}~\bibnamefont
  {Zahedinejad}}, \bibinfo {author} {\bibfnamefont {S.}~\bibnamefont
  {Schirmer}}, \ and\ \bibinfo {author} {\bibfnamefont {B.}~\bibnamefont
  {Sanders}},\ }\href@noop {} {\bibfield  {journal} {\bibinfo  {journal} {Phys.
  Rev. A}\ }\textbf {\bibinfo {volume} {90}},\ \bibinfo {pages} {032310}
  (\bibinfo {year} {2014})}\BibitemShut {NoStop}%
\bibitem [{\citenamefont {August}\ and\ \citenamefont
  {Hern{\'a}ndez-Lobato}(2018)}]{august2018taking}%
  \BibitemOpen
  \bibfield  {author} {\bibinfo {author} {\bibfnamefont {M.}~\bibnamefont
  {August}}\ and\ \bibinfo {author} {\bibfnamefont {J.~M.}\ \bibnamefont
  {Hern{\'a}ndez-Lobato}},\ }\href@noop {} {\bibfield  {journal} {\bibinfo
  {journal} {arXiv preprint arXiv:1802.04063}\ } (\bibinfo {year}
  {2018})}\BibitemShut {NoStop}%
\bibitem [{\citenamefont {Las~Heras}\ \emph {et~al.}(2016)\citenamefont
  {Las~Heras}, \citenamefont {Alvarez-Rodriguez}, \citenamefont {Solano},\ and\
  \citenamefont {Sanz}}]{las2016genetic}%
  \BibitemOpen
  \bibfield  {author} {\bibinfo {author} {\bibfnamefont {U.}~\bibnamefont
  {Las~Heras}}, \bibinfo {author} {\bibfnamefont {U.}~\bibnamefont
  {Alvarez-Rodriguez}}, \bibinfo {author} {\bibfnamefont {E.}~\bibnamefont
  {Solano}}, \ and\ \bibinfo {author} {\bibfnamefont {M.}~\bibnamefont
  {Sanz}},\ }\href@noop {} {\bibfield  {journal} {\bibinfo  {journal} {Phys.
  Rev. Lett.}\ }\textbf {\bibinfo {volume} {116}},\ \bibinfo {pages} {230504}
  (\bibinfo {year} {2016})}\BibitemShut {NoStop}%
\bibitem [{\citenamefont {Banchi}\ \emph {et~al.}(2016)\citenamefont {Banchi},
  \citenamefont {Pancotti},\ and\ \citenamefont {Bose}}]{banchi16learning}%
  \BibitemOpen
  \bibfield  {author} {\bibinfo {author} {\bibfnamefont {L.}~\bibnamefont
  {Banchi}}, \bibinfo {author} {\bibfnamefont {N.}~\bibnamefont {Pancotti}}, \
  and\ \bibinfo {author} {\bibfnamefont {S.}~\bibnamefont {Bose}},\ }\href
  {\doibase 10.1038/npjqi.2016.19} {\bibfield  {journal} {\bibinfo  {journal}
  {npj Quantum Information}\ }\textbf {\bibinfo {volume} {2}} (\bibinfo {year}
  {2016}),\ 10.1038/npjqi.2016.19}\BibitemShut {NoStop}%
\bibitem [{\citenamefont {Sridharan}\ \emph {et~al.}(2008)\citenamefont
  {Sridharan}, \citenamefont {Gu},\ and\ \citenamefont
  {James}}]{sridharan2008gate}%
  \BibitemOpen
  \bibfield  {author} {\bibinfo {author} {\bibfnamefont {S.}~\bibnamefont
  {Sridharan}}, \bibinfo {author} {\bibfnamefont {M.}~\bibnamefont {Gu}}, \
  and\ \bibinfo {author} {\bibfnamefont {M.}~\bibnamefont {James}},\
  }\href@noop {} {\bibfield  {journal} {\bibinfo  {journal} {Phys. Rev. A}\
  }\textbf {\bibinfo {volume} {78}},\ \bibinfo {pages} {052327} (\bibinfo
  {year} {2008})}\BibitemShut {NoStop}%
\bibitem [{\citenamefont {Bukov}\ \emph {et~al.}(2017)\citenamefont {Bukov},
  \citenamefont {Day}, \citenamefont {Sels}, \citenamefont {Weinberg},
  \citenamefont {Polkovnikov},\ and\ \citenamefont {Mehta}}]{bukov2017machine}%
  \BibitemOpen
  \bibfield  {author} {\bibinfo {author} {\bibfnamefont {M.}~\bibnamefont
  {Bukov}}, \bibinfo {author} {\bibfnamefont {A.}~\bibnamefont {Day}}, \bibinfo
  {author} {\bibfnamefont {D.}~\bibnamefont {Sels}}, \bibinfo {author}
  {\bibfnamefont {P.}~\bibnamefont {Weinberg}}, \bibinfo {author}
  {\bibfnamefont {A.}~\bibnamefont {Polkovnikov}}, \ and\ \bibinfo {author}
  {\bibfnamefont {P.}~\bibnamefont {Mehta}},\ }\href@noop {} {\  (\bibinfo
  {year} {2017})},\ \Eprint {http://arxiv.org/abs/1705.00565}
  {arXiv:1705.00565} \BibitemShut {NoStop}%
\bibitem [{\citenamefont {Niu}\ \emph {et~al.}(2018)\citenamefont {Niu},
  \citenamefont {Boixo}, \citenamefont {Smelyanskiy},\ and\ \citenamefont
  {Neven}}]{niu2018universal}%
  \BibitemOpen
  \bibfield  {author} {\bibinfo {author} {\bibfnamefont {M.~Y.}\ \bibnamefont
  {Niu}}, \bibinfo {author} {\bibfnamefont {S.}~\bibnamefont {Boixo}}, \bibinfo
  {author} {\bibfnamefont {V.}~\bibnamefont {Smelyanskiy}}, \ and\ \bibinfo
  {author} {\bibfnamefont {H.}~\bibnamefont {Neven}},\ }\href@noop {}
  {\bibfield  {journal} {\bibinfo  {journal} {arXiv preprint arXiv:1803.01857}\
  } (\bibinfo {year} {2018})}\BibitemShut {NoStop}%
\bibitem [{\citenamefont {August}\ and\ \citenamefont
  {Ni}(2017)}]{august2017using}%
  \BibitemOpen
  \bibfield  {author} {\bibinfo {author} {\bibfnamefont {M.}~\bibnamefont
  {August}}\ and\ \bibinfo {author} {\bibfnamefont {X.}~\bibnamefont {Ni}},\
  }\href@noop {} {\bibfield  {journal} {\bibinfo  {journal} {Physical Review
  A}\ }\textbf {\bibinfo {volume} {95}},\ \bibinfo {pages} {012335} (\bibinfo
  {year} {2017})}\BibitemShut {NoStop}%
\bibitem [{\citenamefont {Swaddle}\ \emph {et~al.}(2017)\citenamefont
  {Swaddle}, \citenamefont {Noakes}, \citenamefont {Smallbone}, \citenamefont
  {Salter},\ and\ \citenamefont {Wang}}]{swaddle17generating}%
  \BibitemOpen
  \bibfield  {author} {\bibinfo {author} {\bibfnamefont {M.}~\bibnamefont
  {Swaddle}}, \bibinfo {author} {\bibfnamefont {L.}~\bibnamefont {Noakes}},
  \bibinfo {author} {\bibfnamefont {H.}~\bibnamefont {Smallbone}}, \bibinfo
  {author} {\bibfnamefont {L.}~\bibnamefont {Salter}}, \ and\ \bibinfo {author}
  {\bibfnamefont {J.}~\bibnamefont {Wang}},\ }\href {\doibase
  10.1016/j.physleta.2017.08.043} {\bibfield  {journal} {\bibinfo  {journal}
  {Phys. Lett. A}\ }\textbf {\bibinfo {volume} {381}},\ \bibinfo {pages} {3391
  } (\bibinfo {year} {2017})}\BibitemShut {NoStop}%
\bibitem [{\citenamefont {F{\"o}sel}\ \emph {et~al.}(2018)\citenamefont
  {F{\"o}sel}, \citenamefont {Tighineanu}, \citenamefont {Weiss},\ and\
  \citenamefont {Marquardt}}]{fosel2018reinforcement}%
  \BibitemOpen
  \bibfield  {author} {\bibinfo {author} {\bibfnamefont {T.}~\bibnamefont
  {F{\"o}sel}}, \bibinfo {author} {\bibfnamefont {P.}~\bibnamefont
  {Tighineanu}}, \bibinfo {author} {\bibfnamefont {T.}~\bibnamefont {Weiss}}, \
  and\ \bibinfo {author} {\bibfnamefont {F.}~\bibnamefont {Marquardt}},\
  }\href@noop {} {\bibfield  {journal} {\bibinfo  {journal} {arXiv preprint
  arXiv:1802.05267}\ } (\bibinfo {year} {2018})}\BibitemShut {NoStop}%
\bibitem [{\citenamefont {Floether}\ \emph {et~al.}(2012)\citenamefont
  {Floether}, \citenamefont {de~Fouquieres},\ and\ \citenamefont
  {Schirmer}}]{floether12robust}%
  \BibitemOpen
  \bibfield  {author} {\bibinfo {author} {\bibfnamefont {F.}~\bibnamefont
  {Floether}}, \bibinfo {author} {\bibfnamefont {P.}~\bibnamefont
  {de~Fouquieres}}, \ and\ \bibinfo {author} {\bibfnamefont {S.}~\bibnamefont
  {Schirmer}},\ }\href {http://stacks.iop.org/1367-2630/14/i=7/a=073023}
  {\bibfield  {journal} {\bibinfo  {journal} {New J. Phys.}\ }\textbf {\bibinfo
  {volume} {14}},\ \bibinfo {pages} {073023} (\bibinfo {year}
  {2012})}\BibitemShut {NoStop}%
\bibitem [{\citenamefont {Bishop}(1995)}]{bishop1995neural}%
  \BibitemOpen
  \bibfield  {author} {\bibinfo {author} {\bibfnamefont {C.}~\bibnamefont
  {Bishop}},\ }\href@noop {} {\emph {\bibinfo {title} {Neural networks for
  pattern recognition}}}\ (\bibinfo  {publisher} {Oxford University Press,
  Oxford, UK},\ \bibinfo {year} {1995})\BibitemShut {NoStop}%
\bibitem [{\citenamefont {Goodfellow}\ \emph {et~al.}(2016)\citenamefont
  {Goodfellow}, \citenamefont {Bengio}, \citenamefont {Courville},\ and\
  \citenamefont {Bengio}}]{goodfellow2016deep}%
  \BibitemOpen
  \bibfield  {author} {\bibinfo {author} {\bibfnamefont {I.}~\bibnamefont
  {Goodfellow}}, \bibinfo {author} {\bibfnamefont {Y.}~\bibnamefont {Bengio}},
  \bibinfo {author} {\bibfnamefont {A.}~\bibnamefont {Courville}}, \ and\
  \bibinfo {author} {\bibfnamefont {Y.}~\bibnamefont {Bengio}},\ }\href@noop {}
  {\emph {\bibinfo {title} {Deep learning}}},\ Vol.~\bibinfo {volume} {1}\
  (\bibinfo  {publisher} {MIT Press, Cambridge, MA, USA},\ \bibinfo {year}
  {2016})\BibitemShut {NoStop}%
\bibitem [{\citenamefont {Koehn}(2009)}]{koehn2009statistical}%
  \BibitemOpen
  \bibfield  {author} {\bibinfo {author} {\bibfnamefont {P.}~\bibnamefont
  {Koehn}},\ }\href@noop {} {\emph {\bibinfo {title} {Statistical machine
  translation}}}\ (\bibinfo  {publisher} {Cambridge University Press,
  Cambridge, UK},\ \bibinfo {year} {2009})\BibitemShut {NoStop}%
\bibitem [{\citenamefont {Bahdanau}\ \emph {et~al.}(2014)\citenamefont
  {Bahdanau}, \citenamefont {Cho},\ and\ \citenamefont
  {Bengio}}]{bahdanau2014neural}%
  \BibitemOpen
  \bibfield  {author} {\bibinfo {author} {\bibfnamefont {D.}~\bibnamefont
  {Bahdanau}}, \bibinfo {author} {\bibfnamefont {K.}~\bibnamefont {Cho}}, \
  and\ \bibinfo {author} {\bibfnamefont {Y.}~\bibnamefont {Bengio}},\
  }\href@noop {} {\  (\bibinfo {year} {2014})},\ \Eprint
  {http://arxiv.org/abs/1409.0473} {arXiv:1409.0473} \BibitemShut {NoStop}%
\bibitem [{\citenamefont {Hochreiter}\ and\ \citenamefont
  {Schmidhuber}(1997)}]{hochreiter1997long}%
  \BibitemOpen
  \bibfield  {author} {\bibinfo {author} {\bibfnamefont {S.}~\bibnamefont
  {Hochreiter}}\ and\ \bibinfo {author} {\bibfnamefont {J.}~\bibnamefont
  {Schmidhuber}},\ }\href {\doibase 10.1162/neco.1997.9.8.1735} {\bibfield
  {journal} {\bibinfo  {journal} {Neural Computation}\ }\textbf {\bibinfo
  {volume} {9}},\ \bibinfo {pages} {1735} (\bibinfo {year} {1997})}\BibitemShut
  {NoStop}%
\bibitem [{\citenamefont {Schuster}\ and\ \citenamefont
  {Paliwal}(1997)}]{schuster1997bidirectional}%
  \BibitemOpen
  \bibfield  {author} {\bibinfo {author} {\bibfnamefont {M.}~\bibnamefont
  {Schuster}}\ and\ \bibinfo {author} {\bibfnamefont {K.~K.}\ \bibnamefont
  {Paliwal}},\ }\href@noop {} {\bibfield  {journal} {\bibinfo  {journal} {IEEE
  Transactions on Signal Processing}\ }\textbf {\bibinfo {volume} {45}},\
  \bibinfo {pages} {2673} (\bibinfo {year} {1997})}\BibitemShut {NoStop}%
\bibitem [{\citenamefont {Graves}\ \emph {et~al.}(2005)\citenamefont {Graves},
  \citenamefont {Fern{\'a}ndez},\ and\ \citenamefont
  {Schmidhuber}}]{graves2005bidirectional}%
  \BibitemOpen
  \bibfield  {author} {\bibinfo {author} {\bibfnamefont {A.}~\bibnamefont
  {Graves}}, \bibinfo {author} {\bibfnamefont {S.}~\bibnamefont
  {Fern{\'a}ndez}}, \ and\ \bibinfo {author} {\bibfnamefont {J.}~\bibnamefont
  {Schmidhuber}},\ }in\ \href@noop {} {\emph {\bibinfo {booktitle}
  {International Conference on Artificial Neural Networks}}}\ (\bibinfo
  {organization} {Springer},\ \bibinfo {year} {2005})\ pp.\ \bibinfo {pages}
  {799--804}\BibitemShut {NoStop}%
\bibitem [{\citenamefont {Abadi}\ \emph {et~al.}(2016)\citenamefont {Abadi},
  \citenamefont {Barham}, \citenamefont {Chen}, \citenamefont {Chen},
  \citenamefont {Davis}, \citenamefont {Dean}, \citenamefont {Devin},
  \citenamefont {Ghemawat}, \citenamefont {Irving}, \citenamefont {Isard},
  \citenamefont {Kudlur}, \citenamefont {Levenberg}, \citenamefont {Monga},
  \citenamefont {Moore}, \citenamefont {Murray}, \citenamefont {Steiner},
  \citenamefont {Tucker}, \citenamefont {Vasudevan}, \citenamefont {Warden},
  \citenamefont {Wicke}, \citenamefont {Yu},\ and\ \citenamefont
  {Zheng}}]{abadi2016tensorflow}%
  \BibitemOpen
  \bibfield  {author} {\bibinfo {author} {\bibfnamefont {M.}~\bibnamefont
  {Abadi}}, \bibinfo {author} {\bibfnamefont {P.}~\bibnamefont {Barham}},
  \bibinfo {author} {\bibfnamefont {J.}~\bibnamefont {Chen}}, \bibinfo {author}
  {\bibfnamefont {Z.}~\bibnamefont {Chen}}, \bibinfo {author} {\bibfnamefont
  {A.}~\bibnamefont {Davis}}, \bibinfo {author} {\bibfnamefont
  {J.}~\bibnamefont {Dean}}, \bibinfo {author} {\bibfnamefont {M.}~\bibnamefont
  {Devin}}, \bibinfo {author} {\bibfnamefont {S.}~\bibnamefont {Ghemawat}},
  \bibinfo {author} {\bibfnamefont {G.}~\bibnamefont {Irving}}, \bibinfo
  {author} {\bibfnamefont {M.}~\bibnamefont {Isard}}, \bibinfo {author}
  {\bibfnamefont {M.}~\bibnamefont {Kudlur}}, \bibinfo {author} {\bibfnamefont
  {J.}~\bibnamefont {Levenberg}}, \bibinfo {author} {\bibfnamefont
  {R.}~\bibnamefont {Monga}}, \bibinfo {author} {\bibfnamefont
  {S.}~\bibnamefont {Moore}}, \bibinfo {author} {\bibfnamefont
  {D.}~\bibnamefont {Murray}}, \bibinfo {author} {\bibfnamefont
  {B.}~\bibnamefont {Steiner}}, \bibinfo {author} {\bibfnamefont
  {P.}~\bibnamefont {Tucker}}, \bibinfo {author} {\bibfnamefont
  {V.}~\bibnamefont {Vasudevan}}, \bibinfo {author} {\bibfnamefont
  {P.}~\bibnamefont {Warden}}, \bibinfo {author} {\bibfnamefont
  {M.}~\bibnamefont {Wicke}}, \bibinfo {author} {\bibfnamefont
  {Y.}~\bibnamefont {Yu}}, \ and\ \bibinfo {author} {\bibfnamefont
  {X.}~\bibnamefont {Zheng}},\ }in\ \href@noop {} {\emph {\bibinfo {booktitle}
  {12th USENIX Symposium on Operating Systems Design and Implementation}}},\
  Vol.~\bibinfo {volume} {16}\ (\bibinfo {year} {2016})\ pp.\ \bibinfo {pages}
  {265--283}\BibitemShut {NoStop}%
\bibitem [{ten()}]{tensorflow}%
  \BibitemOpen
  \href {https://www.tensorflow.org/} {\enquote {\bibinfo {title}
  {{TensorFlow}: An open-source machine learning framework for everyone},}\
  }\BibitemShut {NoStop}%
\bibitem [{\citenamefont {Mezzadri}(2007)}]{mezzadrigenerate}%
  \BibitemOpen
  \bibfield  {author} {\bibinfo {author} {\bibfnamefont {F.}~\bibnamefont
  {Mezzadri}},\ }\href@noop {} {\bibfield  {journal} {\bibinfo  {journal}
  {NOTICES of the AMS}\ }\textbf {\bibinfo {volume} {54}},\ \bibinfo {pages}
  {592} (\bibinfo {year} {2007})}\BibitemShut {NoStop}%
\bibitem [{\citenamefont {Miszczak}(2012)}]{miszczak12generating}%
  \BibitemOpen
  \bibfield  {author} {\bibinfo {author} {\bibfnamefont {J.}~\bibnamefont
  {Miszczak}},\ }\href {\doibase 10.1016/j.cpc.2011.08.002} {\bibfield
  {journal} {\bibinfo  {journal} {Comput. Phys. Commun.}\ }\textbf {\bibinfo
  {volume} {183}},\ \bibinfo {pages} {118} (\bibinfo {year}
  {2012})}\BibitemShut {NoStop}%
\bibitem [{\citenamefont {Banchi}\ \emph {et~al.}(2017)\citenamefont {Banchi},
  \citenamefont {Burgarth},\ and\ \citenamefont
  {Kastoryano}}]{banchi2017driven}%
  \BibitemOpen
  \bibfield  {author} {\bibinfo {author} {\bibfnamefont {L.}~\bibnamefont
  {Banchi}}, \bibinfo {author} {\bibfnamefont {D.}~\bibnamefont {Burgarth}}, \
  and\ \bibinfo {author} {\bibfnamefont {M.~J.}\ \bibnamefont {Kastoryano}},\
  }\href {\doibase 10.1103/PhysRevX.7.041015} {\bibfield  {journal} {\bibinfo
  {journal} {Phys. Rev. X}\ }\textbf {\bibinfo {volume} {7}},\ \bibinfo {pages}
  {041015} (\bibinfo {year} {2017})}\BibitemShut {NoStop}%
\bibitem [{qco()}]{qcontrol_lstm_approx}%
  \BibitemOpen
  \href@noop {} {\enquote {\bibinfo {title} {Approximation of quantum control
  using lstm},}\ }\bibinfo {note}
  {\url{https://github.com/ZKSI/qcontrol_lstm_approx}}\BibitemShut {NoStop}%
\bibitem [{qut(012 )}]{qutip}%
  \BibitemOpen
  \href {http://qutip.org/} {\enquote {\bibinfo {title} {{QuTiP} - {Quantum}
  {Toolbox} in {Python}},}\ } (\bibinfo {year} {2012-})\BibitemShut {NoStop}%
\bibitem [{\citenamefont {Johansson}\ \emph {et~al.}(2012)\citenamefont
  {Johansson}, \citenamefont {Nation},\ and\ \citenamefont {Nori}}]{qutip1}%
  \BibitemOpen
  \bibfield  {author} {\bibinfo {author} {\bibfnamefont {J.}~\bibnamefont
  {Johansson}}, \bibinfo {author} {\bibfnamefont {P.}~\bibnamefont {Nation}}, \
  and\ \bibinfo {author} {\bibfnamefont {F.}~\bibnamefont {Nori}},\ }\href
  {\doibase 10.1016/j.cpc.2012.02.021} {\bibfield  {journal} {\bibinfo
  {journal} {Comput. Phys. Commun.}\ }\textbf {\bibinfo {volume} {183}},\
  \bibinfo {pages} {1760 } (\bibinfo {year} {2012})}\BibitemShut {NoStop}%
\bibitem [{\citenamefont {Johansson}\ \emph {et~al.}(2013)\citenamefont
  {Johansson}, \citenamefont {Nation},\ and\ \citenamefont {Nori}}]{qutip2}%
  \BibitemOpen
  \bibfield  {author} {\bibinfo {author} {\bibfnamefont {J.}~\bibnamefont
  {Johansson}}, \bibinfo {author} {\bibfnamefont {P.}~\bibnamefont {Nation}}, \
  and\ \bibinfo {author} {\bibfnamefont {F.}~\bibnamefont {Nori}},\ }\href
  {\doibase 10.1016/j.cpc.2012.11.019} {\bibfield  {journal} {\bibinfo
  {journal} {Comput. Phys. Commun.}\ }\textbf {\bibinfo {volume} {184}},\
  \bibinfo {pages} {1234 } (\bibinfo {year} {2013})}\BibitemShut {NoStop}%
\end{thebibliography}%
%%%%%%%%%%%%%%%%%%%%%%%%%%%%%%%%%%%%%%%%%%%%%%%%%%%%%%%%%%%%%%%%%%%%%%%%%%%%%%%%
	
\end{document}